\documentclass[10pt,twocolumn,english,aps,prb,reprint]{revtex4}
\usepackage[T1]{fontenc}
\usepackage[latin9]{inputenc}
\usepackage{color}
\usepackage{amsmath}
\usepackage{amssymb}
\usepackage{graphicx}

\makeatletter

\providecommand{\tabularnewline}{\\}

\@ifundefined{textcolor}{}
{%
 \definecolor{BLACK}{gray}{0}
 \definecolor{WHITE}{gray}{1}
 \definecolor{RED}{rgb}{1,0,0}
 \definecolor{GREEN}{rgb}{0,1,0}
 \definecolor{BLUE}{rgb}{0,0,1}
 \definecolor{CYAN}{cmyk}{1,0,0,0}
 \definecolor{MAGENTA}{cmyk}{0,1,0,0}
 \definecolor{YELLOW}{cmyk}{0,0,1,0}
}

\usepackage{babel}

\usepackage{babel}

\usepackage{babel}

\makeatother

\usepackage{babel}
\begin{document}

\title{Enhanced persistent photoconductivity in $\delta$-doped LaAlO$_{3}$/SrTiO$_{3}$
heterostructures}

\author{A. Rastogi$^{1}$, J. J. Pulikkotil$^{2}$, and
R. C. Budhani$^{1,2}$}

\affiliation{$^{1}$Condensed Matter Low Dimensional Systems Laboratory, Department
of Physics, Indian Institute of Technology, Kanpur 208016, India \\
 $^{2}$Division of Quantum Phenomena \& Applications, CSIR-National
Physical Laboratory, Dr. K. S. Krishnan Marg, New Delhi 110012, India}
\email{rcb@nplindia.org}
\begin{abstract}
We report the effect of $\delta$-doping at LaAlO$_{3}$/SrTiO$_{3}$ interface with LaMnO$_{3}$ monolayers on the photoconducting (PC) state. The PC is realized by exposing the samples to broad band optical radiation of a quartz lamp and 325 and 441 nm lines of a He-Cd laser. Along with the significant modification in electrical transport which drives the pure LaAlO$_{3}$/SrTiO$_{3}$ interface from metal-to-insulator with increasing LaMnO$_{3}$ sub-monolayer thickness, we also observe an enhancement in the photo-response and relaxation time constant. Possible scenario for the PC based on defect-clusters, random potential fluctuations and large lattice relaxation models have been discussed. For pure LaAlO$_{3}$/SrTiO$_{3}$, the photoconductivity appears to originate from inter-band transitions between Ti-derived $3d$ bands which are $e_{g}$ in character and O 2p - Ti $t_{2g}$ hybridized bands. The band structure changes significantly when fractional layers of LaMnO$_{3}$ are introduced. Here the Mn $e_{g}$ bands ($\approx1.5$ eV above the Fermi energy) within the photo-conducting gap lead to a reduction in the photo-excitation energy and a gain in overall photoconductivity.
\end{abstract}
\maketitle

\section{introduction}
The origin of the two-dimensional electron gas (2DEG) at the interface of artificially tailored oxide heterostructures \cite{Muller,Ohtomo,Hwang,Caviglia,Thiel} is attributed to both intrinsic and extrinsic factors. While the former is accounted for the polar catastrophe model, \cite{Huijben,Millis} extrinsic factors are associated with defects, such as oxygen vacancies \cite{Siemons,Kalabukhov,Herranz1} and inter-site cation mixing \cite{Tokura,Pulikkotil} that are introduced in the system during film growth. \cite{Huijben1,Herranz} The latter argument is substantiated by the observation of a clear dependence of the interface conductivity on pressure during the film growth. \cite{Huijben,Herranz} Furthermore, the observation of persistent photo-conductivity (PPC) with large relaxation time also points to the role of defect induced states in these oxide heterostructures. \cite{Rastogi,Rastogi01,Rastogi02} However, there also exists conclusive evidence that defects in the form of O-vacancies in the SrTiO$_{3}$ substrate are not the only responsible factor that leads to 2DEG at the interface. Had so been the case, then irrespective of the nature of the films, a 2DEG would have been observed for most of the perovskite oxides deposited on SrTiO$_{3}$, including for LaCrO$_{3}$ and LaMnO$_{3}$. Beyond, we note that doping induces electronic phase transition and metallicity in both LaCrO$_{3}$ and LaMnO$_{3}$.\cite{Sakai,Schiffer,Park} On the other hand, transport measurements find that the LaCrO$_{3}$ and LaMnO$_{3}$ films deposited on TiO$_{2}$ terminated SrTiO$_{3}$ substrate results in no 2DEG at the interface, irrespective of the film thickness and deposition conditions. \cite{Chambers01,Barriocanal}

In general, a 2DEG state is observed at the interface of a non-polar
(TiO$_{2}$ terminated SrTiO$_{3}$) and a polar material (such as
LaAlO$_{3}$) motif. Under these circumstances, the in-built electric
field at the interface causes bending of conduction and valence band
edges of SrTiO$_{3}$, leading to a triangular potential well which
is filled by electrons transferred from the LaAlO$_{3}$ over-layers.
\cite{Nakagawa,Stengel,Pentcheva,Chen,Son,Ramesh} Although, this
model explains many of the experimental observations, it fails to
account for the observed charge carrier density that has its dependence
on the thickness of the deposited LaAlO$_{3}$ film. \cite{Thiel,Caviglia}
Beyond, the model also has little support from the photo-emission
spectroscopy studies. \cite{Segal,Chambers,Janicka,Yoshimatsu}

A fundamental understanding of the underlying mechanism for the formation
of the 2DEG therefore requires use of perturbative techniques, which
will lead to injection of additional charge carriers at the interface
and selective and controlled modification of the interface chemistry.
While the former is realized by electrostatic gating and photo-excitation,
\cite{Thiel,Rastogi,Rastogi02} the latter can be achieved by $\delta$-
doping at the interface, a concept commonly used in III-V compound
semiconductor quantum wells. \cite{Tusi}

The effect of electrostatic gating of LaAlO$_{3}$/SrTiO$_{3}$ interface
has been studied extensively. The gate field either draws charges
towards the interface from the over-layers or pushes charge towards
it resulting in a gate controlled metal-insulator phase transition.
\cite{Thiel,Caviglia01} It has also been shown earlier that the
electronic transport in oxide heterostructures can be altered significantly
on exposure to electromagnetic radiation of optical frequency making
these potential candidates in opto-electronic applications. \cite{Rastogi01,Rastogi02,Rastogi}
For the conducting interface of LaTiO$_{3}$/SrTiO$_{3}$ and LaAlO$_{3}$/SrTiO$_{3}$,
the PPC is seen in the spectral range of
300 $-$ 400 nm. It is also noticed that the magnitude and relaxation
dynamics of the photo-conducting state depends on the growth temperature
of LaTiO$_{3}$ and LaAlO$_{3}$, but is independent of the number
of the film overlayers, since the response is only energetically close
to the band gap of SrTiO$_{3}$. It has been previously shown that SrTiO$_3$ (single crystal and thin films) displays wide range of properties such as ferroelectric, \cite{Ishikawa,Tikhomirov} persistence photoconductivity \cite{Feng} etc. Therefore, it is quite evident that the SrTiO$_{3}$ substrate plays a crucial role in determining the
electronic and optical properties in these heterostructures. The
PPC seen in oxide interfaces is quite similar
to that reported in the III-V semiconductor heterostructures. \cite{Arslan,Stormer}
In analogy with the latter systems, the slow decay of the photo-current
in oxides can be associated with the defect induced states in the
SrTiO$_{3}$ substrate. The longer life-time of the photo-induced
carriers is largely attributed to the less dispersive Ti $3d$ bands,
which make up the conduction band \cite{Rastogi02} in these systems.

We have observed a systematic metal-to-insulator transition in LaAlO$_{3}$/SrTiO$_{3}$ system by controlled
$\delta$-doping at the interface with LaMnO$_{3}$ monolayers. \cite{Rastogi03}
Our present experiment rules out the possibility of interfacial intermixing
as the cause of metallic conduction in LaAlO$_{3}$/SrTiO$_{3}$ system,
because such a reaction of La/Sr at the interface would lead to formation
of La$_{1-x}$Sr$_{x}$TiO$_{3}$ which should be conducting based
on the simple valence argument. Our experiment also rules out the
reduction of SrTiO$_{3}$ as the cause of 2DEG, as reduction would
take place even in the case of $\delta$-LaMnO$_{3}$ monolayer deposition,
resulting in interfacial conductivity irrespective of the Mn concentration
at the interface. While the present experiment validates the polarization
catastrophe argument for the 2DEG formation, it remains to be seen
what would happen to photo-conductivity when transfer of electrons
from LaAlO$_{3}$ overlayers are inhibited by the $\delta$-LaMnO$_{3}$
monolayer.

Here we present the results of photo-conductivity measurements on
the $\delta$-doped LaAlO$_{3}$/SrTiO$_{3}$ heterostructures, and compared them with the photo-response of the undoped LaAlO$_{3}$/SrTiO$_{3}$
system. The $\delta$-doping is accomplished by means of a selective
deposition of fractional monolayers of LaMnO$_{3}$ on the TiO$_{2}$
terminated SrTiO$_{3}$ surface, followed by layer-by-layer growth
of the LaAlO$_{3}$ over the LaMnO$_{3}$ layer. It has been shown
previously that LaMnO$_{3}$ films on SrTiO$_{3}$ remain non-conducting
irrespective of their thickness. \cite{Barriocanal} Therefore, a
controlled deposition of Mn ions at the LaAlO$_{3}$/SrTiO$_{3}$
interface is expected to provide a method to control the electronic phases
and phase transition. In accordance, we find that the resistivity
of the $\delta$-doped LaAlO$_{3}$/SrTiO$_{3}$ increases non-trivially
as the $\delta$-LaMnO$_{3}$ monolayer at the interface is thickened.
Beyond, the results of photo-conductivity measurements reported here
also include the following: (i) A larger photo-response with increasing
$\delta$-doped LaMnO$_{3}$ interfacial layer thickness and, (ii)
persistent photo-conductivity following a stretched exponential behavior
with its decay constant increasing in proportion with the LaMnO$_{3}$
monolayer thickness. Explanation to these results are sought in terms
of a few well established models and also partly from the band structure
calculations.

\section{Experimental details}

The samples were grown by Pulsed Laser Deposition at 800$^{o}$C and
the thickness of LaAlO$_{3}$ deposited on TiO$_{2}$ terminated ($001$)
SrTiO$_{3}$ was kept at 20 unit-cell (uc). The deposition procedure, conditions
and growth parameters are described in our earlier reports. \cite{Rastogi01,Rastogi02,Rastogi03}
The $\delta$-doping of the interface has been achieved by controlled
growth ($\simeq$ $0.012$ - $0.015$ nm/s) of a LaMnO$_{3}$ sub-monolayer
prior to commencing deposition of LaAlO$_{3}$ film. In its bulk form, LaMnO$_{3}$
is a strongly correlated anti-ferromagnetic insulator. A schematic
of the interfacial $\delta$-doping with LaMnO$_{3}$ is shown in
the inset of Fig.\ref{Fig01}. To perform the transport measurements,
Ag/Cr electrodes were deposited on the films (drain and source) and
back of the substrate (gate) by thermal evaporation using shadow masks.
Linear current versus voltage characteristics between drain and source
confirm the Ohmic behavior of the electrodes, while the leakage current
between the gate and source was less than $10$ nA. The photo-excitation
experiments were carried out in a closed cycle helium optical cryostat where
the samples were exposed to the broad band radiation of a Xenon
lamp, whose spectrum contains $\simeq$ $3.5$\% of UV radiation,
and two single wavelength (441 and 325 nm) lines of a He-Cd laser through a quartz window.

\section{Results}

\subsection{Electrical Resistivity}

\begin{figure}[h]
\includegraphics[scale=0.32]{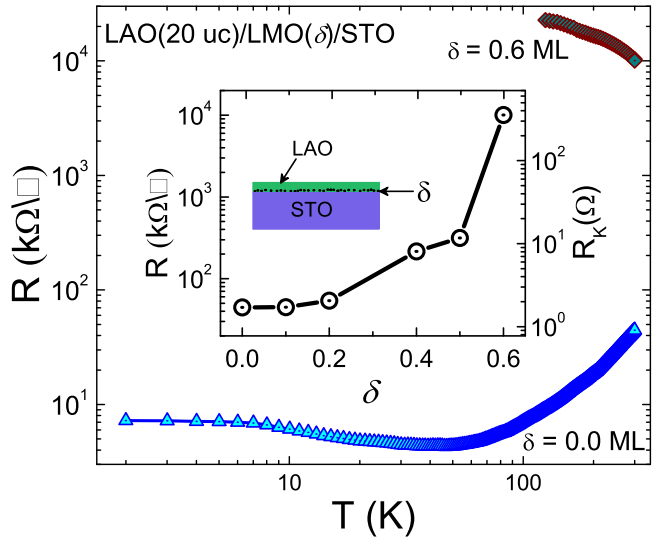}

\caption{\label{Fig01} (Color Online) The temperature dependence of the sheet
resistance ($R_{\square}$) for the samples with and without LaMnO$_{3}$
($\delta$= 0 and $0.6$ ML) at the LaAlO$_{3}$/SrTiO$_{3}$ interface.
The LaAlO$_{3}$/SrTiO$_{3}$ sample shows a metallic behavior as
the temperature is lowered to $\simeq$ 30 K followed by an upturn,
while the sample with $0.6$ ML LaMnO$_{3}$ at the interface shows
insulating behavior. \cite{Rastogi03} The resistance goes beyond
the measurement limit below $\simeq$$120$ K. The inset shows the
variation of resistance at 300 K with the LaMnO$_{3}$ layer thickness in two
different units; with the left axis representing it in units of k$\Omega$
while other axis is represented in the units of quantum resistance
R$_{k}$ = h/e$^{2}$, where h and e are the Planck's constant and
electronic charge.}
\end{figure}

The temperature dependence of the sheet resistance ($R_{\square}$)
of the LaAlO$_{3}$/SrTiO$_{3}$ and that of the $\delta$-LaMnO$_{3}$
doped samples is shown in Fig.\ref{Fig01}. Consistent with the previous
reports, a $20$ unit cell thick LaAlO$_{3}$ film on TiO$_{2}$ terminated
SrTiO$_{3}$ substrate showed metallic characteristics. \cite{Rastogi01,Rastogi02,Rastogi03}
However, on embedding LaMnO$_{3}$ sub-monolayer at the interface, we
observe a systematic transition to insulating state. We have also
noticed that the critical thickness of the LaAlO$_{3}$ over-layer
required to induce 2DEG in $\delta$-LaMnO$_{3}$ doped heterostructures
is proportional to the sub-monolayer thickness of the $\delta$-LaMnO$_{3}$.
\cite{Rastogi03} This study also revealed that $\simeq0.6$ uc thick
LaMnO$_{3}$ drives the interface to insulating state when the
LaAlO$_{3}$ overlayer thickness was of 20 uc. We note that the critical
thickness of LaMnO$_{3}$ to make the interface insulating is much smaller when LaAlO$_{3}$ layer
is only 10 uc. The plot of $R_{\square}$(T) for $\delta$-doped LaMnO$_{3}$
with sub-monolayer thickness is shown in Fig.\ref{Fig01} along with
$R_{\square}$(T) of undoped LaAlO$_{3}$/SrTiO$_{3}$.

\subsection{Photo-response}

\begin{figure}[!]
\includegraphics[width=8cm,height=15cm]{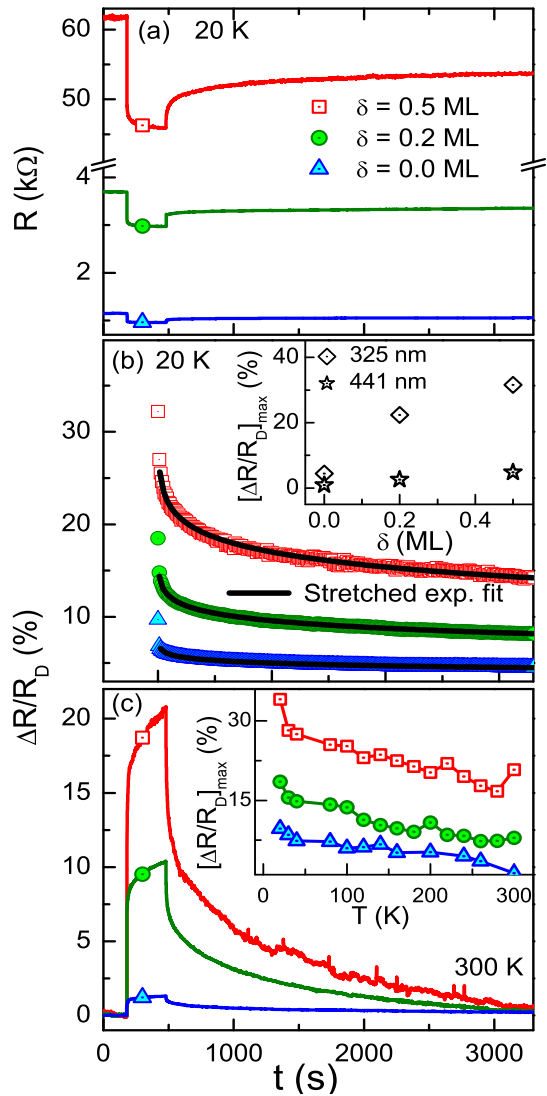}

\caption{\label{Fig02} (Color Online) (a) shows the change in the channel
resistance at 20 K as a function $\delta$ layer thickness. The main
panel (b) and (c) respectively show the relaxation of the normalized
resistance for different $\delta$-doping at 20 and 300 K after switching
off the illumination from a halogen lamp. The recovery dynamics follow
a stretched exponential behavior which are represented as solid lines
in (b). The inset of (b) shows the relative change in the resistance
at $300$ K upon radiating the samples with 325 and 441 nm lines of
a He-Cd laser. A comparison of the photo-response as a function of
temperature for different samples is made in inset (c). The $0.5$
monolayer LaMnO$_{3}$ shows a three fold increase in the photo-response
in comparison with the $\delta$ = 0 sample.}
\end{figure}

Fig. \ref{Fig02}(a) shows the time evolution of resistance of
three samples with $\delta\simeq0,0.2$ and $0.5$ ML during the period
when sample was exposed to the light and then allowed to recover in
dark at 20 K. To better understand the recovery process, we have defined
a normalized resistance $\Delta$R/R$_{D}$, where $\Delta$R = R(t)-R$_{D}$
with R(t) being the resistance at time t and R$_{D}$ the resistance
in dark. The details of these measurement have been described in an
earlier report. \cite{Rastogi02} Fig. \ref{Fig02}(b $\&$ c) reveal
that the relative change in the resistance on photo-exposure increases
with $\delta$-LaMnO$_{3}$ layer thickness, while the recovery process
slows down. It is important to point out here that the most change
in the resistance on photo-exposure is triggered by the ultra-violet
(UV) component of the quartz halogen lamp radiation. This fact is
further established when we expose the sample to a monochromatic radiations
($\lambda$ = 325 and 441 nm) of He-Cd laser. These results are shown
in the inset of Fig. \ref{Fig02}(b), from which we find that relative
change in the resistance of all heterostructures under consideration
is significantly suppressed for 441 nm radiation, in comparison to
that observed under the 325 nm radiation. In Table \ref{table}, we
list the values of $\Delta$R/R$_{D}$ for all three samples achieved
under 325 nm, 441 nm and xenon lamp radiation. The comparative study
as highlighted in Table \ref{table} shows that the overall photo-response
increases with increasing $\delta$ doping of LaMnO$_{3}$ at the
interface. Also, our data suggests that the threshold wavelength to
induce photo-conductivity in these heterostructures shift towards
higher wavelengths with $\delta$ doping. This effect has been qualitatively understood
by means of band structure calculations, which is discussed in the next section.

\begin{table}[h]
\begin{tabular}{l|c|c|c}
\hline
 & $\delta$ = 0  & $\delta$ = 0.2  & $\delta$ = 0.5 \tabularnewline
\hline
\multicolumn{1}{l|}{($\Delta$R/R$_{D}$)$_{325nm}$ } & 0.0446  & 0.2236  & 0.3156 \tabularnewline
\multicolumn{1}{l|}{($\Delta$R/R$_{D}$)$_{441nm}$ } & 0.0094  & 0.0258  & 0.0481 \tabularnewline
\multicolumn{1}{l|}{($\Delta$R/R$_{D}$)$_{Broadband}$ } & 0.0187  & 0.1079  & 0.2089 \tabularnewline
\hline
\end{tabular}\caption{ \label{table}A comparison of relative change in the resistance using
different source of radiation.}
\end{table}

\begin{figure}[h]
\includegraphics[scale=0.33]{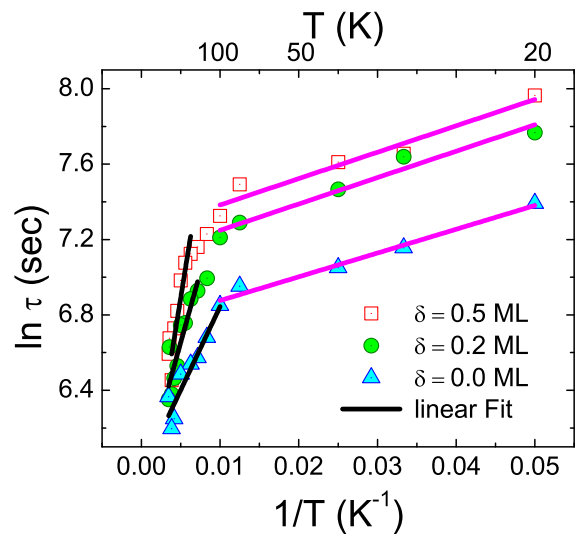}

\caption{\label{Fig03} (Color Online) Comparison of the relaxation time constant
$\left(\tau\right)$ of three system as indicated by the legends.
The solid curves are the fits to the Arrhenius equation. The relaxation
time constant scales in proportion with increasing LaMnO$_{3}$ sub-monolayer
thickness. For system corresponding to $0.5$ ML thick LaMnO$_{3}$,
the relaxation time constant was estimated to be higher.}
\end{figure}

In Fig. \ref{Fig02}(c) we show the $\Delta$R/R$_{D}$ data taken
at 300 K and the temperature dependence of maximum $\Delta$R/R$_{D}$
(just after closing the shutter). The response decreases monotonically
with increasing temperature and it is higher in $\delta$-doped samples.
In general, the recovery to the resistive state on shutting off the
light is well described by a stretched exponential of the form R $\propto$
$\exp\left[-\left(\frac{t}{\tau}\right)^{\beta}\right]$ with $0$$\leq$$\beta$
$\leq$ $1$, and, $\tau$ being the relaxation time constant. A fit
(shown by the solid lines in Fig. \ref{Fig02}(b)) yields $\tau$ varying
from $1700$ to $3000$ seconds as the thickness of the $\delta$-layer
is increased to $0.5$ uc. Correspondingly, the exponent $\beta$
goes from $0.2$ to $0.8$. Further, the recovery dynamics depend
strongly on temperature. In order to understand the recovery process
and the mechanisms involved, we fit the data to Arrhenius equation
given as $\tau$ = $\tau_{0}$$\exp\left(-\frac{\Delta U}{k_{B}T}\right)$,
where $\Delta U$ and $k_{B}$ are the activation energy for detrapping
the photo-generated carriers and Boltzmann constant, respectively.
The plots of $\ln\left(\tau\right)$ against $\left(\frac{1}{T}\right)$
are shown in Fig.\ref{Fig03}. Clearly, two distinct temperature regions
of activation can be seen in the Arrhenius plot. At low temperatures
T $<$ $100$ K, the calculated value of $\Delta U$ is in the range
$1$ $-$ $2$ meV, for all samples. However, at higher temperatures,
the $\Delta U$ for $0.5$ uc thick LaMnO$_{3}$ is estimated as $\simeq$
$22$ meV, which is almost three times higher in comparison to the
activation energy of LaAlO$_{3}$/SrTiO$_{3}$ system ($\simeq$ $8$
meV).

A sudden change in the slope about $100$ K is observed in Arrhenous plot (Fig. \ref{Fig03}). This temperature is very close to the structural phase transition of SrTiO$_3$ ($\sim$ $105$ K). We believe that such a phase transition would be of lesser significance, as the cubic to tetragonal transition of SrTiO$_3$ is partly due to its latent paraelectric nature, which is intimately associated with the d$^0$-ness of the Ti ions in SrTiO$_3$. In this case, the interface modifies the occupation of Ti with extra electrons being doped and therefore such a transition is not expected in the SrTiO$_3$ motif near to the interface, although one may expect the same to happen deep down in the substrate.

\begin{figure}[h]
\includegraphics[scale=0.3]{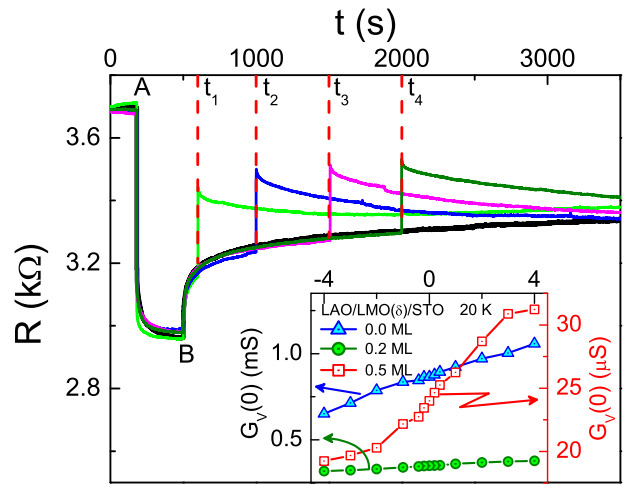}

\caption{\label{Fig04} (Color Online) Main panel shows the effect of electric field in the PPC recovery state for $\delta$ = $0.2$ ML sample at 20 K. Points `A' and `B' in the figure show the time at which the illumination is turned `ON' and `OFF' respectively. This photo-induced recovery state is then subjected to the gate field (E$_{g}$) at times t1, t2, t3, and t4. Inset shows the zero bias conductance (G$_{V}$(0)) of all three samples in dark at 20 K in the range $\pm$4 kV/cm.}
\end{figure}

The electrical conductivity of these structures undergoes strong modulation
when a gate field is applied. It has been shown earlier that the electrostatic
and photon fields act on two different sets of charge carriers. \cite{Rastogi02}
While the recovery time from the photo-illuminated state stretches
over several hours, the system was found to recover to the unperturbed
normal state with in microseconds after switching off the gate field.
Hence, the role of $\delta$-layer influencing the migration of these
two sets of carriers needs to be examined. We have studied the influence
of electric field perturbation on the photo-induced relaxation processes
of the $\delta$-doped LaMnO$_{3}$ interfaces in comparison with
that of the pure LaAlO$_{3}$/SrTiO$_{3}$ system.

In Fig.\ref{Fig04} we illustrate our measurement scheme to study
the effects of a gate electric field on the recovery process. At point `A'
of the plot in Fig. \ref{Fig04} the sample is exposed to light, which
triggers a sharp drop in resistance followed by saturation. At point
`B' the illumination is turned off and the recovery dynamics thereafter
is examined by applying a gate field (E$_{g}$) of $-2$kV/cm after a certain time gap, which has been changed from t$_1$ to t$_4$ in four sets of exposure-recovery experiments. We find that the resistance changes spontaneously
on gating. While positive E$_{g}$ enhances the channel conductivity,
negative E$_{g}$ tend to reduce it. For better understanding of the
effect of gate field alone, we have measured source to drain I-V characteristic
of these samples under different gate fields. The slope of these linear
curves at the origin gives the zero-bias conductance G$_{V}$(0) which
is plotted in the inset of Fig. \ref{Fig04}. We observe an order in
magnitude decreased in the G$_{V}$($0$), in the $\delta=0.5$ ML
sample.

\subsection{Electronic structure}

The threshold energy required to impart photo-conductivity in LaAlO$_{3}$/SrTiO$_{3}$
corresponds to that of near ultra-violet wavelength. Numerical calculations
based on density functional theory are consistent with this picture.
\cite{Rastogi02} The optical transitions are associated to those
states which are in the range $\pm2$ eV above and below the Fermi
energy, thereby providing the estimate of photo-conducting threshold
wavelength as $<500$ nm.

\begin{figure}
\includegraphics[width=8cm,height=9cm]{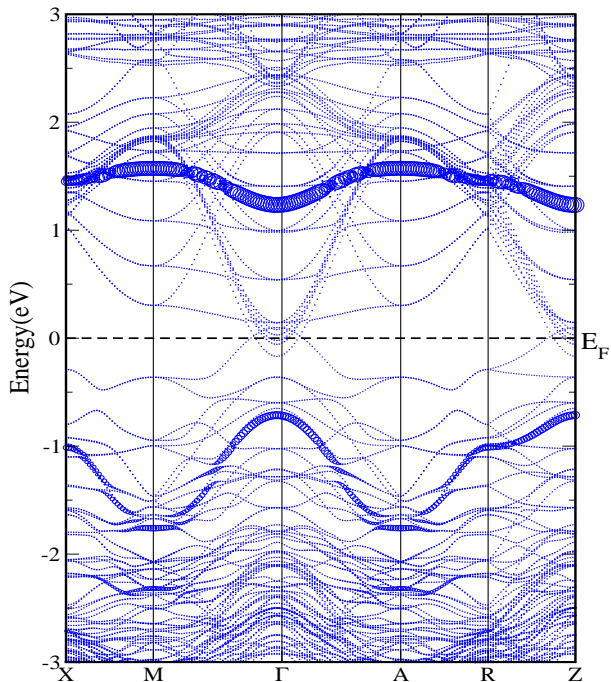}

\caption{\label{BNDS} Fat band representation of the Mn $3d$ majority states (circle) of the $\delta$-doped LaMnO$_{3}$ in LaAlO$_{3}$/SrTiO$_{3}$ heterostructures,
calculated using the LDA+U method, with U= $8$ eV. The conduction
band around E $\simeq$$1.5$ eV is derived of the Mn $3d_{z^{2}}$
spin up states, while those in the valence band in the range $1.7$
$<$ E (eV) $<$ $0.5$, corresponds to the Mn $3d_{xy}$ spin up
states.}
\end{figure}

However, the photo-response of the $\delta$-doped LaAlO$_{3}$/SrTiO$_{3}$
heterostructures show subtle signatures of the photo-response threshold
shifting to lower photon energy, which suggests a modification in the band structure caused by
the Mn ions in the vicinity of the interface. In Fig.\ref{BNDS},
we show the band structure of the $\delta$-doped LaAlO$_{3}$/SrTiO$_{3}$.
A fully relaxed super-cell of 2$\times$2$\times$9 dimension with
$4.5$ unit-cells of SrTiO$_{3}$ as substrate with TiO$_{2}$ termination
is modeled in the present study. The $\delta$-doping with $0.5$
ML thick LaMnO$_{3}$ is modeled by Mn-Al chemical disorder
near to the interface. The Brillouin zone integration were carried
out on a $11$$\times$$11$$\times$$2$ $k-$space grid with the
WIEN2K \cite{Blaha} parameters $RK_{max}$ and $G_{max}$ to be as
$7$ and $24$, respectively. The numerical details of the structure
optimization, self-consistent total energy and that of the electronic
structure are similar to those described in Ref. 17.
Both in local density approximation (LDA) and generalized gradient
approximation (GGA), Mn ions modeled at the interface of LaAlO$_{3}$/SrTiO$_{3}$
system yield a metallic state, with relatively high density of states,
with the prominent states being the Mn $3d$ states. It is well known
that such spurious states at the Fermi energy are due consequence
of the missing Coulomb correlation term in the Hamiltonian. In the
present case, the Coulomb correlation effects (U$_{eff}$) of the
Mn $3d$ electrons are taken into consideration by the LDA+U$_{eff}$
Hamiltonian, with U$_{eff}$= $8$ eV.

Before we discuss the band structure of the $\delta$-doped LaMnO$_{3}$,
recall from Fig.$9$ of Ref. 17 that the calculated
optical conductivity of the LaAlO$_{3}$/SrTiO$_{3}$ heterostructure
matches well with that of the experiments. The consistency indicates
the reliability of the band structure generated in the density functional
theoretical formalism. The upper valence band of the LaAlO$_{3}$/SrTiO$_{3}$
system are primarily composed of O $2p$ states of the AlO$_{2}$
layers of the LaAlO$_{3}$ film. Electronic states, $\simeq$ $2$
eV below the Fermi energy, are the O $2p$ states of the TiO$_{2}$
layers of the substrate. On the other hand, the states above Fermi
energy are primarily composed of Ti $3d$ orbitals. At energies $\geq$
2 eV above Fermi energy, one finds a bunching of Ti $3d$ bands of
primarily of the $e_{g}$ character. These states are relatively localized
over a very narrow energy interval. Thus, photo-conductivity in LaAlO$_{3}$/SrTiO$_{3}$
heterostructures results from the inter-band transitions between Ti
derived $e_{g}$ bands and that of the O $2p$ - Ti $t_{2g}$ amalgamation
at the Fermi energy.

In case of $\delta$-doped LaAlO$_{3}$/SrTiO$_{3}$ with LaMnO$_{3}$
monolayers, the band structure as shown in Fig.\ref{BNDS} shows
localized Mn $3d_{z^{2}}$ states (e$_{g}$ states) positioned $\simeq$$1.5$
eV above Fermi energy. The corresponding Mn $3d_{xy}$ states ($t_{2g}$)
in the valence band are relatively more dispersed over a wider energy
range. Thus, it is evident that the positioning of the Mn $e_{g}$
states within the photo-conducting gap leads to a decrease in the
photo-excitation energy, and therefore is expected to increase the
overall photo-response. Thus, for small wavelength induced excitations
one find the transition to occur between the states that lie between
$\pm$$2$eV of the Fermi energy, while the small photo-generation
of carriers at the $441$ nm is  associated with the Ti $3d_{xy}$
states near the Fermi energy and the Mn $e_{g}$ bands which are relatively
$1.5$ eV above the Fermi energy.

The computed band structure also qualitatively explains the relatively
slow relaxation of the photo-generated carriers to the normal state
in the $\delta$-doped systems with that of the pure LaAlO$_{3}$/SrTiO$_{3}$.
The energy-time relation of the uncertainty principle suggests that
the life-time of photo-generated carriers in a given band would be
inversely proportional to the energy band width. Along these perspectives,
finding that the Mn $e_{g}$ bands which are highly localized and
positioned at $1.5$ eV above the Fermi energy would serve as localized
traps for the photo-induced charge carriers, thereby increasing the
relaxation time in $\delta$-doped systems in comparison with the
parent system.

\section{Discussion}

For abrupt interfaces of oxide heterostructures, the polar catastrophe
model is quite robust to account for the origin of 2DEG. The model
asserts on band bending effects via electronic reconstruction due
to the in-built electric field at the interface of the polar (LaAlO$_{3}$)$-$nonpolar
(SrTiO$_{3}$) materials. PPC in these
systems can therefore be related to the model proposed by Queisser
and Theodorou (QT), where its emergence is associated with the macroscopic
potential barriers induced by the band bending effects. \cite{Queisser}
Such potential barriers lead to the PPC effect by spatial separation
of the photo-generated electrons and holes. With one type of carrier
being trapped, the other remains free and causes excess conductivity.
\cite{Lee} The model can be verified by studying the excitation-energy
dependence of PPC. For this we used the two excitation energies afforded
from He-Cd laser, which gives one of the excitation energy (2.8 eV)
less while the other one larger (3.8 eV) than the band gap energy
of SrTiO$_{3}$. The results are displayed in Table- \ref{table} and inset
of Fig. \ref{Fig02}(b). For pure LaAlO$_{3}$/SrTiO$_{3}$ heterostructures,
we observed very small change in the resistance and also weak signature
of PPC with the photo-excitation below the absorption edge, while
for $\delta$-doped samples a large PPC effect could be seen, which
increases monotonically with the $\delta$-layer thickness. These
results infer that the PPC of undoped sample have little contribution
arising from spatial separation of the photo-generated carries, primarily
due to the band bending effects. While in the case of $\delta$-doped
systems, this effect seems to be less prominent. The change in the
resistance of $\delta$-doped systems with lower excitation can be
attributed to the finite absorption of incident illumination by in-gap
states present in the SrTiO$_{3}$ band gap and reduce the possibility of band
bending scenario. It has also been shown the 2DEG confinement can
be explained by the formation of metal induced gap states in the band
gap of SrTiO$_{3}$ rather than the band bending \cite{Janicka} and
no measurable band bending is observed in La$_{1-x}$Al$_{1+x}$O$_{3}$/SrTiO$_{3}$(001)
hetero-junctions. \cite{Qiao} Thus, the model based on the spatial
separation of the photoexcited electrons and holes by macroscopic
potential barrier due to band bending looks less appropriate to account
for the PPC here. However, if the band bending is due to the O-vacancies,
then the associated barrier would spatially separate the electrons
that constitute the 2DEG. Although, this can lead to some justification
of the QT model to the observed PPC effects in the present hetero-structured
systems, the observed decay kinetics does not seem to support the
model as a predominant process for inducing PPC. The above picture
of band bending is further established by the analysis of the PPC
decay dynamics as proposed in QT model for artificially tailored heterostructures.
\cite{Queisser} We tried to fit the decay to a logarithmic behavior
of the form, $\Delta R/R_{D}=a-b*ln(1+t/\tau)$, where a and b are
constants, with respect to time. We find that the logarithmic fit
becomes poorer and deviates significantly from the measured data as
the $\delta$-layer thickness is increased. However, the undoped system
still displays a better $\chi^{2}$ tolerance to the fit and low residual
values as compared to the $\delta$-doped samples, which strengthen
our argument.

Various other models have also been proposed to account for the PPC
effects in solid state materials. \cite{Lang,Queisser,Mooney,Jiang,Iseler,Schubert}
Earliest of all was associated to inhomogeneities in the samples,
caused either by bonding configurations, spatial variation in the
chemical composition and/or in local charge densities. A much more
advanced model was due to Theodorou and Symeonides, \cite{Theodorou}
who proposed the role of defect clusters in the sample. Such defects
may be inevitable and are generally introduced during the growth itself.
On preliminary grounds, the defect-cluster model may look appropriate
to account for PPC in these heterostructures, as the synthesis technique
employed in depositing $\delta$-doped LaAlO$_{3}$ films on SrTiO$_{3}$
substrates are by means of the laser ablation. This highly energetic
deposition process could result in clusterization of the Mn and Al
ions forming a disordered LaAl$_{1-x}$Mn$_{x}$O$_{3}$ monolayer.
In cases where the local inhomogeneities lead to a density of the
defects which is larger than the carrier density, a macroscopic potential
barrier may be formed. Such defect-clusters may host traps, and upon
electron capture would become charged. In turn, this would induce
a charge of opposite polarity around the cluster, thus spatially separating
the electron-holes pairs, impeding their recombination and resulting
in PPC. \cite{Theodorou,Theodorou1} However, the defect-cluster model
predicts larger PPC in proportion to the thickness of the conducting
layer. The latter, however, is in contrast with our observation in
$\delta$-doped LaMnO$_{3}$ in the LaAlO$_{3}$/SrTiO$_{3}$ interface.
We find that as the $\delta$-doped sub-monolayer thickness is increased
(larger Al/Mn disorder), the 2DEG conductivity decreases, but with
an enhancement in the PPC. Thus, the defect-cluster model also seems
less appropriate to account for the larger PPC seen in the $\delta$-doped
LaMnO$_{3}$ in the samples.

A relevant model that appears to describe the PPC in the present case
is based on potential fluctuations in the material. \cite{Permogorov,Cohen,Jiang,Jiang1,Jiang2,Lin}
Here the spatial separation between photo-excited charge carriers
by random local-potential fluctuations induced by compositional variations
is held responsible for PPC. \cite{Jiang} Within the framework of
the polar catastrophe model, the origin of 2DEG is due to electronic
reconstruction of the Ti ions at the interface. Thus, the 2DEG wave-function
would be thought as a composition of Ti$^{3+}$ and Ti$^{4+}$ states.
Moreover, the Mn ions in the monolayer would also exhibit multiple
valence states. Certainly therefore, an energy barrier emanating from
such a charge fluctuation would also act as a source of potential
fluctuations. Besides, there is also a likelihood of chemical disorder
due to random distribution of Al and Mn ions in the delta doped layer.
Therefore, both chemical and valence state disorder can induce an
uneven potential landscape at the interface of LaAlO$_{3}$/SrTiO$_{3}$ with $\delta$-doping
of the interface. However, this model predicts PPC to be observed
at high temperatures, with a well defined transition temperature,
which is in contrast with our observations, where enhanced PPC in
seen at lower temperatures. But, we note that the decay kinetics predicted
by the random potential fluctuation model is similar to that of a
stretched exponential function.

The lattice relaxation model \cite{Lang,Lang1,Zhou} is one of the
most widely accepted description for the PPC effects in semiconductor
heterostructures. The model stipulates photo-excitation of carriers
from defect induced deep-level traps . A barrier is created by lattice
relaxation thereby preventing the recapture of mobile carriers. The
source of the lattice relaxation could be ionic mismatch and/or vacancies.
However, in the present context, mismatch in the lattice constants
and/or that in the constituents ionic-radii, does not seems to be
relevant. Therefore, the structural relaxation would be mainly due
to oxygen vacancies, and hence these defects render a negative-U center.
At low temperatures, thermal energy is not sufficient enough to overcome
the barrier, and thus the photo-generated carriers remain in these
shallow states, resulting in PPC. The PPC buildup and decay kinetics,
in general, fit to a stretched exponential curve, which is similar
to our observations for the oxide heterostructures. \cite{Rastogi01,Rastogi02}
Due to the inherent propensity of LaMnO$_{3}$ to attract oxygen,
its presence in the vicinity of the interface can induce more vacancies
in the substrate leading to a non-equilibrium chemical configuration.
Thus, while the photo-ionization of oxygen specific defects in the
quantum well region appears to be the dominant mechanism for PPC in
oxide heterostructures, one cannot rule out the contribution of the
spatially separated charges across the interface facilitated by the
factors that have been mentioned above.

Moreover, as both Mn and Al occupy the substitutional body-centered
site of the perovskite structure, the role of Mn ions on the lattice
can also be partly inferred from the PPC data. It has been argued
in the case of III-V and II-IV semiconductor heterostructures, and
also in oxide heterostructures that the 2DEG at the interface is
derived out of lattice strain. Within the realms of alloy theory,
the solubility of Mn ions in LaAlO$_{3}$ matrix, infers to a lowering
of lattice strain . In general, reduction of strain decrease the activation
energy with Mn at\% in the monolayer, which, however, is in contradiction
with our observation. Thus, the increase in activation energy with
$\delta$-doping at the interface may be attributed to the enhanced
electron-electron interaction between the electrons in the trap and
the unpaired electrons of the Mn ion. Such an event lead to a narrowing
of the optical trap potential, which would narrow with increasing
Mn at\% at the interface, leading to a higher repulsive barrier for
the electrons to return to the oxygen mediated defect centers, after
photo-ionization. However, we note that there does not exist to the
best of our knowledge, a scheme to estimate the strength of Coulomb
scattering from the presence of the deep traps in the quantum well
and its effect on transport properties. However, it is reported that
in ZnSe/(Zn,Cd,Mn)Se heterostructured systems, the 2DEG exhibits a
strong PPC that becomes more pronounced when the Mn$^{2+}$ concentration
is increased. \cite{Ray} The enhanced PPC in this system is also
accompanied by a decrease in sample mobility, suggesting a connection
between the deep traps introduced by the magnetic Mn ions and enhanced
Coulomb scattering.

\section{conclusion}

In summary, we have investigated the effect of $\delta$-doping on
the photo-response of LaAlO$_{3}$/SrTiO$_{3}$ interfaces. The photo-response
of the pure LaAlO$_{3}$/SrTiO$_{3}$ heterostructure is found to
be sensitive to near ultra-violet radiation which shifts towards the
lower photon energy upon doping the interface with LaMnO$_{3}$. The
doped samples show relatively large photo-response and time constant
of recovery in comparison to the undoped sample. Based on theoretical
calculations, we establish that the slow relaxation arises due to
localized Mn $e_{g}$ states (3$d_{z^{2}}$) which are situated $\sim$
1.5 eV above the Fermi level. The positioning of these Mn $3d$ $e_{g}$
bands within the photo-conducting gap shifts the photo-response threshold
to higher photon wavelengths with increasing $\delta$-doping in the
heterostructures. We also have made an attempt to understand the decay
dynamics of the photo-conducting state . Our experimental findings
demonstrate that the defect-cluster and random fluctuation models
are less appropriate to describe the large photo-response and high
values of activation energies for the recovery seen in $\delta$-doped
samples. On other hand, the lattice relaxation model is found to be
in better agreement. Moreover, our arguments also suggest that the
increase in the activation energy could be attributed to the strong
electron-electron interaction of the Mn-ions in the $\delta$-doped
monolayer at the interface.
\begin{acknowledgments}
The authors gratefully acknowledge discussions with S.
Auluck. A.R. would like to acknowledge the Council of Scientific and
Industrial Research (CSIR), India and Indian Institute of Technology
Kanpur for financial support. R.C.B. acknowledges financial support
from J. C. Bose National Fellowship of the Department of Science and
Technology, Government of India. \end{acknowledgments}

\end{document}